\documentclass[accepted]{melba}

\usepackage{mwe} 

\usepackage{amsmath,amsfonts}
\usepackage{ulem}
\usepackage[capitalize]{cleveref}
\usepackage{adjustbox} 
\usepackage{booktabs} 
\usepackage{multirow}
\usepackage{natbib}
\usepackage{stfloats}
\usepackage{tabularx}
\newcolumntype{L}[1]{>{\raggedright\arraybackslash}p{#1}} 

\melbaid{2025:030}  
\doi{10.59275/j.melba.2025-c586}
\melbaauthors{Bercea, Cattin, Schnabel, and  Wolleb}  
\email{julia.wolleb@yale.edu}
\volume{3}
\firstpageno{687}  
\melbayear{2025}  
\datesubmitted{2025-03-31}  
\datepublished{2025-11-27}  

\ShortHeadings{Denoising Diffusion Models for Anomaly Localization// in Medical Images}{Bercea, Cattin, Schnabel, and Wolleb}
\ShortHeadings{Denoising Diffusion Models for Anomaly Localization in Medical Images}{Bercea, Cattin, Schnabel, and Wolleb}

\title{Denoising Diffusion Models for Anomaly Localization \\ in Medical Images}

\author{
	\firstname Cosmin I. \surname Bercea\aff{1,2} \orcid{0000-0003-2628-2766},
	\name Philippe C. \surname Cattin\aff{3} \orcid{0000-0001-8785-2713},
    \name Julia A. \surname Schnabel\aff{1,2,4} \orcid{0000-0001-6107-3009},
    \name Julia \surname Wolleb\aff{3,5,6} \orcid{0000-0003-4087-5920 }
}

\affiliations{
	\num 1 \addr School of Computation, Information and Technology, Technical University of Munich, Germany \\
	\num 2 \addr Institute of Machine Learning in Biomedical Imaging, Helmholtz Munich, Germany\\
	\num 3 \addr Department of Biomedical Engineering, University of Basel, Allschwil, Switzerland \\
    \num 4 \addr School of Biomedical Engineering and Imaging Sciences, King's College London, United Kingdom \\
    \num 5 \addr Department of Biomedical Informatics and Data Science, Yale University School of Medicine, New Haven, CT, USA\\
    \num 6 \addr Yale Biomedical Imaging Institute, Yale University, New Haven, CT, USA
}

\abstract{
	This review explores anomaly localization in medical images using denoising diffusion models. After providing a brief methodological background of these models, including their application to image reconstruction and their conditioning using guidance mechanisms, we provide an overview of available datasets and evaluation metrics suitable for their application to anomaly localization in medical images. In this context, we discuss supervision schemes ranging from fully supervised segmentation to semi-supervised, weakly supervised, self-supervised, and unsupervised methods, and provide insights into the effectiveness and limitations of these approaches. Furthermore, we highlight open challenges in anomaly localization, including detection bias, domain shift, computational cost, and model interpretability. Our goal is to provide an overview of the current state of the art in the field, outline research gaps, and highlight the potential of diffusion models for robust anomaly localization in medical images.}


\begin{document}

\twocolumn[\maketitle]


\section{Introduction}

\label{sec:1}

\begin{figure}[b!]
\centering
\includegraphics[width=\linewidth]{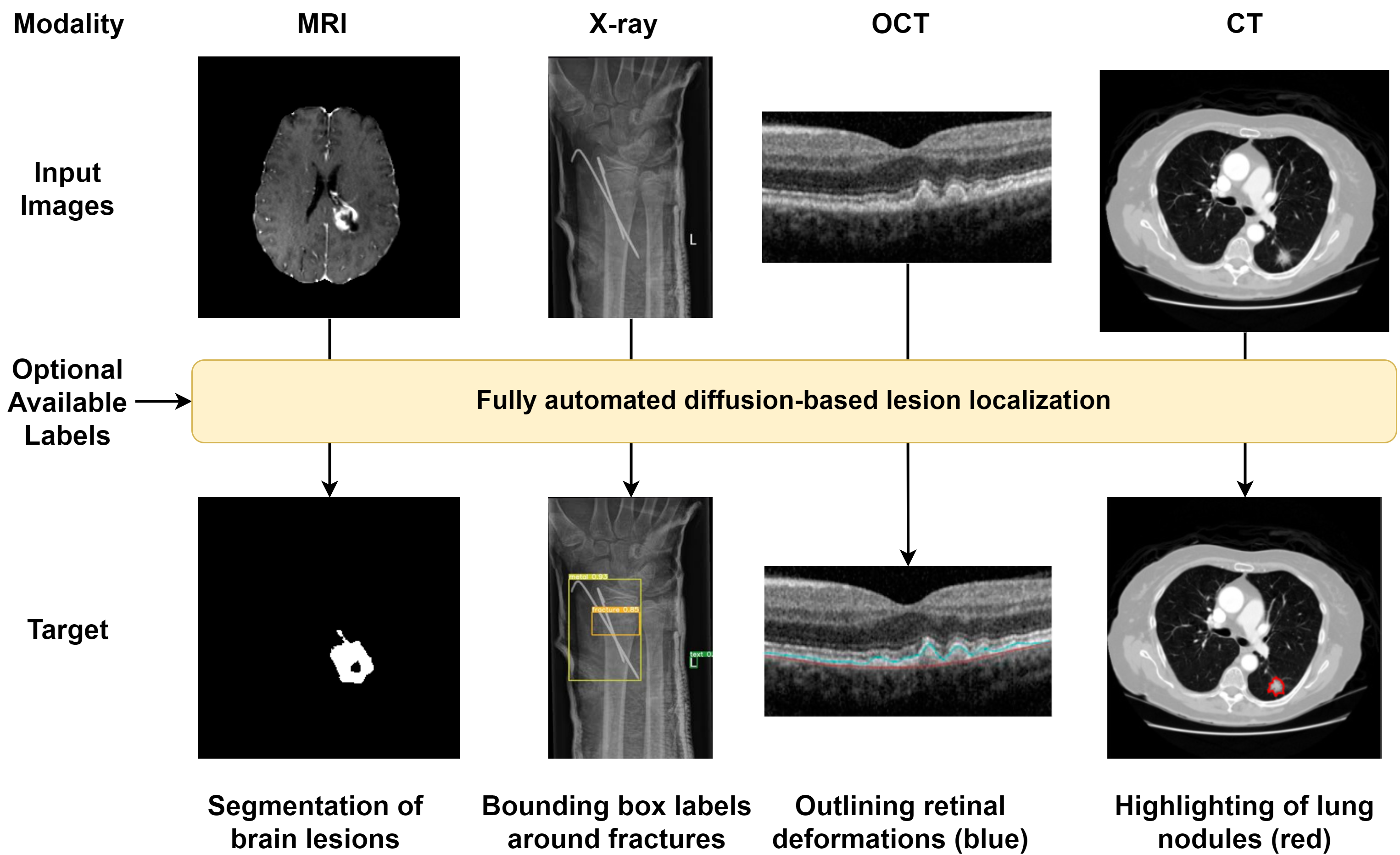}
\caption{Anomaly localization is the task of outlining pathological regions in medical images. In this work, we present an overview of automated solutions based on denoising diffusion models. }
\label{fig:overview_anomalytypes}       
\end{figure}

Anomaly localization in medical images refers to the process of identifying abnormal areas or regions within images of various modalities, such as X-ray, computed tomography (CT), magnetic resonance imaging (MRI), or optical coherence tomography (OCT), see~\cref{fig:overview_anomalytypes}.

These anomalies may be indicative of a range of conditions or diseases, including tumors, fractures, organ malformations, and vascular abnormalities. Precisely locating and highlighting these pathological changes assists medical experts in diagnosing and monitoring diseases. However, manual analysis is a tedious task requiring significant time and expertise for accurate interpretation. Deep learning algorithms have played a key role in the automatic processing of medical images~\citep{chen2022recent,zhou2023deep}. Specifically, fully supervised segmentation algorithms like U-Nets \citep{ronneberger2015u} have been instrumental in automatic lesion segmentation, providing detailed delineation of abnormal areas. In addition, generative models such as Variational Autoencoders (VAEs) \citep{kingma2019introduction}  and Generative Adversarial Networks (GANs) \citep{goodfellow2020generative} have shown promising results in anomaly localization, even with limited labeled data.
In recent years, denoising diffusion models have emerged as a class of deep generative models that estimate complex probability distributions~\citep{ho2020denoising}. They have gained attention for their ability to generate high-quality samples, outperforming GANs in image synthesis \citep{dhariwal2021diffusion}. 
While diffusion models have primarily been proposed for image generation, their ability to generate healthcare data has enabled real-world applications that address data scarcity and privacy concerns \citep{giuffre2023harnessing}. In addition, they have been adapted to data augmentation and to solve downstream tasks like denoising, inpainting, and image restoration. In this review, we focus on the methodological aspects of denoising diffusion models (DDMs) for lesion localization, a direction with strong potential to support radiologists in the future \citep{prasad2024revolutionizing}.
Regulatory frameworks are still evolving, with consensus guidelines introduced in \cite{lekadir2025future} and the EU Artificial Intelligence Act \citep{edwards2021eu} representing the first comprehensive legislation regulating AI across the European Union.

In this work, we investigate the potential and limitations of diffusion models for anomaly localization in medical imaging under various data and label availability scenarios. 
Anomaly segmentation refers to delineating pathological regions in an image, which is a pixel-level classification task that aims to precisely trace the extent of abnormalities. This fully supervised approach typically assumes an anomaly is present and focuses on where and how it manifests.
In contrast, anomaly detection determines whether an anomaly exists, and can be applied at the image, bounding-box, or pixel level.
In this review, we use the term “anomaly localization” to span both segmentation and detection. Following \cite{baugher1986boundary}, localization refers to estimating the approximate location of an anomaly, while delineation involves precisely tracing its boundaries. Our use of “localization” thus covers a range of approaches depending on label availability and model design — from coarse localization to fine-grained segmentation.

This review aims to provide a comprehensive overview of the current state of the art, identify research gaps, and highlight open challenges. For a more detailed comparison of denoising diffusion models to GANs and VAEs, we refer to related work by \cite{bercea2025evaluating} and \cite{friedrich2024deep}.
We begin with a brief methodological background on diffusion models, emphasizing their application in image reconstruction and conditioning through guidance mechanisms. We then present an overview of available datasets and evaluation metrics tailored for anomaly localization in medical images. Next, we explore various supervision schemes, ranging from fully supervised segmentation to semi-supervised, weakly supervised, self-supervised, and unsupervised methods, and provide insights into the effectiveness and limitations of diffusion models in these scenarios, as well as their generalizability to unknown or rare diseases. Finally, we address open challenges, such as detection bias, domain shift, clinical validation, computational cost, and model interpretability.

\section{Background}
\label{sec:2}

In this section, we first provide a brief theoretical background on the basic framework of denoising diffusion models. In~\cref{subsection: basicidea}, we explain how diffusion models can be applied to reconstruction-based anomaly detection. Finally, in~\cref{subsec:evolution}, we present the various guidance mechanisms necessary for conditioning diffusion models for medical anomaly localization.

\subsection{Denoising Diffusion Models}
\label{subsection:theory}
Denoising diffusion models are a class of generative models inspired by thermodynamic processes \citep{sohl2015deep}.  The core idea is to gradually corrupt an input signal into pure noise using a forward diffusion process, then learn to reverse this process step by step. This learned reverse process transforms random noise into samples that resemble the original data distribution.
As proposed in \cite{ho2020denoising} and \cite{improving}, the core idea of denoising diffusion models is that for many timesteps $T$, Gaussian noise is added to an input image $x$. This results in a series of noisy images ${x_0, x_1, ..., x_T}$, where the noise level is steadily increased from $0$ (no noise) to $T$ (maximum noise). 
The forward noising process $q$ with variances  $\beta_{1},...,\beta_{T}$ is defined by
\begin{equation}\label{eq:forward}
q(x_{t}|x_{t-1}):=\mathcal{N}(\sqrt{1-\beta _{t}}x_{t-1},\beta _{t}\mathbf{I}).
\end{equation}
This recursion can be written explicitly as 
\begin{equation}\label{eq:property}
x_{t}=\sqrt[]{\bar{\alpha} _{t}}x_{0}+\sqrt[]{1-\bar{\alpha} _{t}}\epsilon, \quad \mbox{with } \epsilon \sim \mathcal{N}(0,\mathbf{I}),
\end{equation}
 $\alpha_{t}:=1-\beta _{t}$ and $\bar{\alpha}_{t}:=\prod_{s=1}^t \alpha _{s}$.
\\
The mathemtical principles, including appropriate choices of the hyperparameters $T$ and $\beta_t$, are thoroughly derived by \cite{ho2020denoising}, \cite{song2020denoising}, and \cite{dhariwal2021diffusion}.
The denoising process $p_{\theta}$ is learned by a deep learning model that predicts $x_{t-1}$ from $x_t$ for any step $t \in \{1,...,T\}$. It is given by
\begin{equation}\label{eq:reverse}
p_{\theta}(x_{t-1}\vert x_t):= \mathcal{N}\bigl(\mu_{\theta}(x_t, t), \Sigma_\theta(x_t,t)\bigr),
\end{equation} 
where $\mu_{\theta}$ and $\Sigma_\theta$ are learned by a time-conditioned neural network following a U-Net architecture \citep{ronneberger2015u}. A memory-efficient U-Net implementation was presented by \cite{bieder2024memory} . In an alternative transformer-based architecture, Diffusion Transformers replace the commonly-used  convolutional U-Net backbone with a transformer that operates on latent patches \citep{peebles2023scalable}. 
By using the reparametrization trick presented by \cite{kingma2013auto}, the output of the U-Net $\epsilon_{\theta}$ is a noise prediction, and the MSE loss used for training simplifies to
\begin{equation}\label{eq4}
\mathcal{L}:= ||\epsilon-\epsilon_{\theta}(\sqrt[]{\bar{\alpha} _{t}}x_{0}+\sqrt[]{1-\bar{\alpha} _{t}}\epsilon, t)||^2_2 , \quad \mbox{with } \epsilon \sim \mathcal{N}(0,\mathbf{I}).
\end{equation}

The mathemtical principles, including appropriate choices of the hyperparameters $T$ and $\beta_t$, are thoroughly derived by \cite{ho2020denoising}, \cite{song2020denoising}, and \cite{dhariwal2021diffusion}

For image generation, we start from $x_T \sim \mathcal{N}(0,\mathbf{I})$ and iteratively go through the denoising process by predicting $x_{t-1}$ for $t \in \{T,...,1\}$.
This sampling procedure of denoising diffusion models can be divided into denoising diffusion probabilistic models (DDPMs) and denoising diffusion implicit models (DDIMs). Training as described in~\cref{eq4} remains the same for both approaches.\\
\\
\textbf{DDPM sampling scheme}: 
As shown in \citep{song2020denoising}, we use the DDPM formulation to predict $x_{t-1}$ from $x_t$ with
\begin{align}\label{eq:sampling}
x_{t-1} = &\sqrt{\bar{\alpha}_{t-1}}\left(\frac{x_t-\sqrt{1-\bar{\alpha}_t}\epsilon_\theta(x_t,t)}{\sqrt{\bar{\alpha}_t}}\right)\\ \nonumber
&+\sqrt{1-\bar{\alpha}_{t-1}-\sigma_t^2}\epsilon_{\theta}(x_t,t)+\sigma_t\epsilon, 
\end{align}
where we choose $\sigma_t = \sqrt{(1 - \bar{\alpha}_{t-1}) / (1 - \bar{\alpha}_t)} \sqrt{1 - \bar{\alpha}_t / \bar{\alpha}_{t-1}}$, and $\epsilon~\sim~\mathcal{N}(0,\mathbf{I})$.  DDPMs thereby have a stochastic element $\epsilon$ in each sampling step.\\
\\
\textbf{DDIM sampling scheme}: In denoising diffusion implicit models, we set ${\sigma_t=0}$ in,~\cref{eq:sampling} , which results in a deterministic sampling process. Then, as derived in \citep{song2020denoising},~\cref{eq:sampling} can be viewed as the Euler method to solve an ordinary differential equation (ODE).\\
\\
\textbf{DDIM noise encoding}:
 We can reverse the generation process by using the reversed ODE of the DDIM sampling scheme. Using enough discretization steps, we can encode $x_{t+1}$ given $x_t$ with
\begin{align}\label{eq:ddim_reversed}
x_{t+1}  =  x_{t}+\sqrt{\bar{\alpha}_{t+1}} \Biggl[ &\left( \sqrt{\frac{1}{\bar{\alpha}_{t}}} - \sqrt{\frac{1}{\bar{\alpha}_{t+1}}}\right) x_t \nonumber \\
+ &\left(\sqrt{\frac{1}{\bar{\alpha}_{t+1}} - 1} - \sqrt{\frac{1}{\bar{\alpha}_{t} }- 1} \right) \epsilon_\theta(x_t,t) \Biggr].
\end{align}
 We denote~\cref{eq:ddim_reversed} as the DDIM noise encoding scheme. By applying~\cref{eq:ddim_reversed} for $t \in \{0,...,T-1\}$, we can encode an image $x_0$ in a noisy image $x_T$. Then,  we recover the identical $x_0$ from $x_T$ by using~\cref{eq:sampling} with $\sigma_t=0$ for $t \in \{T,...,1\}$. \\
 \\
Both the DDPM and DDIM processes are designed to generate synthetic data from random noise, i.e., it is a purely generative model. To adapt this model class to medical downstream task, conditioning mechanisms will be necessary to tailor the model output to the task at hand. This will be discussed in~\cref{subsec:evolution}.

\subsection{Reconstruction-based Anomaly Localization}
\label{subsection: basicidea}
\begin{figure*}[h]
\centering
\includegraphics[width=0.8\linewidth]{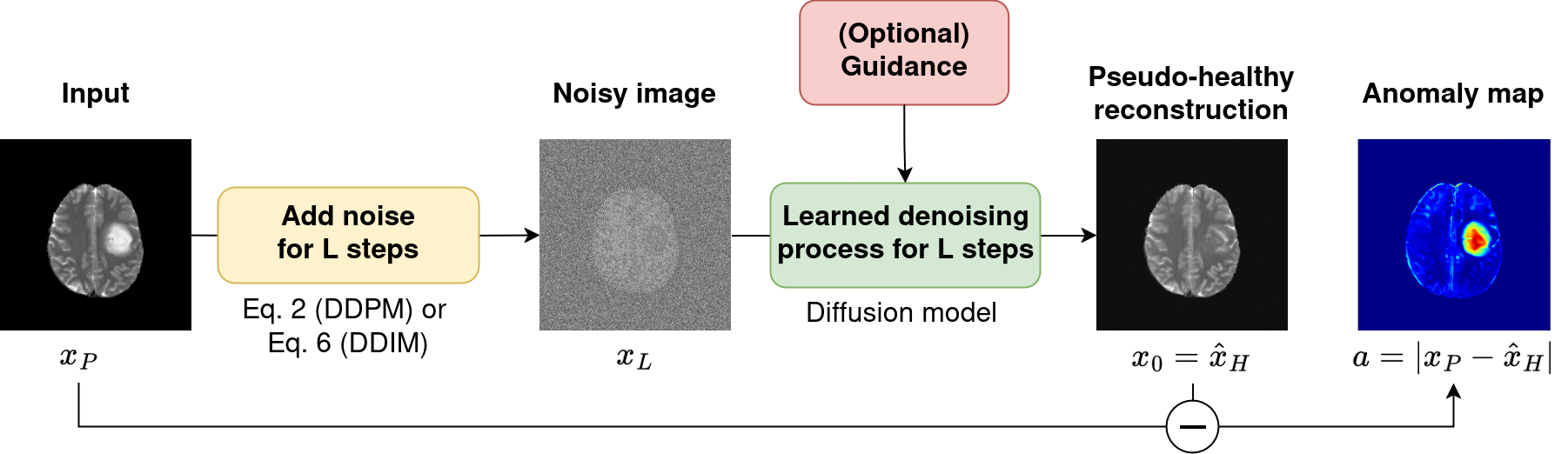}
\caption{In reconstruction-based anomaly localization using diffusion models, we first add noise to an input image $x_P$. During denoising, we translate an input image to its pseudo-healthy counterpart $\hat{x}_H$. The pixel-wise anomaly map is given by the difference between input $x_P$ and output $\hat{x}_H$.}
\label{fig:reconstruction}       
\end{figure*}
\noindent We can leverage image-to-image translation to restore a pseudo-healthy image $\hat{x}_H$ from an input image $x_P$ potentially showing a pathology, as presented in~\cref{fig:reconstruction}. In this work, we consider diffusion-based approaches, where $L$ steps of noise are added to an input image $x_P$, following~\cref{eq:property} or \cref{eq:ddim_reversed} for the DDPM and DDIM schemes, respectively. The hyperparameter $L$ indicates the level of noise added, as well as the number of denoising step taken during the denoising process.
We emphasize the importance of this parameter $L$, as it directly governs how much the denoising process can alter the reconstruction, thereby affecting both sensitivity and specificity; its impact is systematically analyzed by \cite{bercea2023mask} and  \cite{wolleb2022anomaly}.
A diffusion model is trained or guided to generate healthy samples. Various techniques discussed in~ \cref{section4} ensure that only pathological regions are altered while subject-specific anatomical structures are preserved. Finally, by taking the difference between an input image $x_P$ and its pseudo-healthy reconstruction $\hat{x}_H$, a pixel-wise anomaly map $a=|x_P-\hat{x}_H|$ can be defined.

\subsection{The Evolution of Guidance\label{subsec:evolution}}

While the denoising diffusion models described in~\cref{subsection:theory} are designed for unconditional image generation, many medical downstream tasks require adherence to specific requirements, such as being conditioned on an input image or conforming to the characteristics of a particular disease. In addition to scalar conditioning demonstrated in~\citep{nichol2021improved}, more advanced conditioning mechanisms can be explored:

\begin{itemize}

 \item  \textbf{Conditioning through Concatenation:}
\cite{saharia2022palette} proposed \textit{Palette}, where conditioning through concatenation involves appending the conditioning image (e.g., a grayscale or corrupted version) as extra input channels to the noisy image at each denoising step. This allows the model to consistently use spatial guidance from the conditioning input throughout the reverse diffusion process. The approach is simple, stable to train, and broadly applicable to paired image-to-image translation tasks like inpainting, super-resolution, and colorization.

\item \textbf{Gradient Guidance:} 
An external guidance scheme, as presented in \citep{dhariwal2021diffusion}, involves training a classification network $C$ to distinguish between class labels. This network $C(x_t,t)$ shares the encoder architecture of the diffusion model and is trained on noisy input images $x_t$ obtained via~\cref{eq:property}. During sampling, following the DDIM scheme, at each denoising step, $\epsilon_{\theta}$ in~\cref{eq:sampling} is modified to $\hat \epsilon_{\theta} = \epsilon_{\theta}(x_t,t) - s\* \sqrt{1-\bar{\alpha}_t}  \nabla_{x_t} \log C(c|x_t,t)$, with the hyperparameter $s$ scaling the guidance. This steers image generation towards the desired class $c$. The diffusion model $\epsilon_{\theta}$ and the classification model $C$ are trained separately, offering the flexibility to combine gradients from different networks in a plug-and-play fashion without having to retrain the diffusion model \citep{wolleb2022swiss}. However, training two separate networks can be unstable, and output quality depends on the performance of the classification network, introducing potential bias.

\item \textbf{Classifier-free guidance:} 
As proposed in \citep{ho2022classifier}, guidance can be incorporated directly into the latent space of the diffusion model by training a class-conditional diffusion model $\epsilon_{\theta}(x_t,t,c)$. This model can omit class information by selecting $c= \emptyset$ with a certain probability, allowing it to train on both conditional and unconditional objectives by randomly dropping $c$ during training. This method provides robust gradients and the efficiency of training a single model. However, it requires retraining the model for each new classification task.

\item \textbf{Implicit Guidance:}
Guidance can be provided by the input image itself. Patch-based methods \citep{behrendt2024patched} condition the model on input image patches while predicting masked areas. AutoDDPMs \citep{bercea2023mask} further employ masking to remove highly probable anomalous tissues, stitching to integrate healthy parts with pseudo-normal complements, and resampling from their joint noised distribution. In THOR \citep{bercea2024diffusion}, the backward process is guided through intermediate anomaly maps to project pseudo-normal images closer to the inputs. Bernoulli diffusion models \citep{wolleb2024binary} follow the idea of binary latent diffusion models and introduce a masking algorithm based on anomaly scores that can directly be extracted from the model output.

\end{itemize}
A summary of these approaches is provided in~\cref{tab:keyfeatures}.  

\begin{table*}[ht]
\footnotesize
\centering
\renewcommand{\arraystretch}{1.3}
\caption{Comparison of guidance strategies in diffusion models.}  \label{tab:keyfeatures}
\begin{tabularx}{\textwidth}{L{3cm} L{4.8cm} L{2.8cm} L{3cm} L{3.2cm}} 
\toprule

\textbf{Key Feature} & \textbf{Description} & \textbf{Examples} & \textbf{Advantages} & \textbf{Disadvantages} \\
\midrule

Guidance through Concatenation & 
The conditioning on the input image is performed through concatenation at every step of the denoising process. & \cite{saharia2022palette,amit2021segdiff,durrer2023diffusion}
 & 
Stable to train; can be applied to any paired image-to-image translation task & 
Requires paired images \\
\midrule
Gradient Guidance & 
A separate task-specific model is trained on the dataset. During the denoising process, the gradient of this model is used to update the denoising process, guiding it towards desired characteristics. & \cite{dhariwal2021diffusion, wolleb2022anomaly}
 & 
The two models are independent and can be flexibly used in a plug-and-play manner & 
Sampling can be unstable and heavily relies on hyperparameters such as noise level and guidance scale \\
\midrule
Classifier-free Guidance & 
The diffusion model is trained using an optional class label as input in every block of the U-Net. During sampling, the desired class label can be provided to guide generation. & 
\cite{sanchez2022healthy,liu2024controllable} & 
Only one model to train; stable gradients & 
The model needs to be retrained for each new task \\
\midrule
Implicit Guidance & 
Guidance is provided by the input image itself through self-supervised training. & 
\cite{bercea2023mask,wyatt_anodppm, behrendt2024patched} & 
No external conditioning labels required & 
The chosen self-supervision scheme must be carefully adapted to the task \\
\bottomrule
\end{tabularx}

\end{table*}
\section{Datasets}
\begin{table*}[t!]
    \centering 
    \caption{Datasets used for diffusion-based medical anomaly detection, covering various anatomies, modalities, and anomaly types. The table also lists the size of each dataset, the type of annotations provided, and the methods applied.\label{tab:datasets}}
    \setlength{\tabcolsep}{2pt}
    \begin{adjustbox}{width=\linewidth,center} 
        \begin{tabular}{c | c | c | c | c | c | c}
        \toprule
        Anatomy & Dataset & Size & Modality & Anomaly type & Annotation Type & Methods \\ \midrule
        \multirow{15}{*}{Brain} & CROMIS~\cite{Wilson2018} & 311 volumes & CT & Stroke & Pixel-wise & ~\cite{pinaya2022fast}\\
         & KCH/CHRONIC~\cite{Mah2020} & 1704 volumes & CT& Stroke & Pixel-wise & ~\cite{pinaya2022fast}\\ 
         & \multirow{4}{*}{BraTS~\cite{brats1,brats2,brats3}} & \multirow{4}{*}{1251 volumes} & \multirow{4}{*}{Multi-Modal MRI} & \multirow{4}{*}{Tumors} & \multirow{4}{*}{Pixel-wise} & ~\cite{behrendt2024patched} \\
         & & & & & & ~\cite{marimont2023disyre}  \\
         & & & & & & ~\cite{wolleb2022anomaly}  \\
        & & & & & & ~\cite{wolleb2024binary}  \\

        & \textit{Private Clinic} & 22 volumes & T1w MRI & Tumors & Pixel-wise & ~\cite{wyatt_anodppm} \\

        & \multirow{3}{*}{ATLAS~\cite{atlas2022}} & \multirow{3}{*}{655 volumes} & \multirow{3}{*}{Multi-Modal MRI} & \multirow{3}{*}{Stroke} & \multirow{3}{*}{Pixel-wise} & ~\cite{bercea2023mask} \\
         & & & & & & ~\cite{bercea2024diffusion}  \\
         & & & & & & ~\cite{marimont2023disyre}  \\
        & \multirow{2}{*}{WMH~\cite{WMH}} & \multirow{2}{*}{60 volumes} & \multirow{2}{*}{T1w + FLAIR MRI} & White matter & \multirow{2}{*}{Pixel-wise} & \multirow{2}{*}{~\cite{pinaya2022fast}}\\
        & & & & Hyper-intensities & & \\ 
        & \multirow{2}{*}{MSLub~\cite{mslub}} & \multirow{2}{*}{30 volumes} & \multirow{2}{*}{Multi-Modal MRI} & \multirow{2}{*}{MS lesions} & \multirow{2}{*}{Pixel-wise} & ~\cite{behrendt2024patched}\\
         & & & & & & ~\cite{pinaya2022fast}  \\
        & FastMRI+~\cite{zbontar2018fastmri,zhao2021fastmri+} & 1001 volumes & Multi-Modal MRI & 30 anomaly types & Bounding Boxes & ~\cite{bercea2025evaluating}\\  
         \midrule
        
        Chest & Chexpert~\cite{irvin2019chexpert} & 224316 scans & X-ray & 14 anomaly types & Bounding boxes & ~\cite{wolleb2022anomaly} \\ \midrule 
        \multirow{2}{*}{Retina} & \multirow{2}{*}{OCT2017~\cite{Kermany2018LabeledOC}} & 30K healthy & \multirow{2}{*}{OCT} & \multirow{2}{*}{3 anomaly types} & \multirow{2}{*}{Image-wise} & \multirow{2}{*}{~\cite{wolleb2024binary}}  \\ & & 3000 pathological& &&\\  \midrule 
        \multirow{1}{*}{Wrist} & \multirow{1}{*}{GRAZPED~\cite{nagy2022pediatric}} & 20327 scans & \multirow{1}{*}{X-ray} & \multirow{1}{*}{8 anomaly types} & \multirow{1}{*}{Bounding boxes} & \multirow{1}{*}{~\cite{bercea2024diffusion}}  \\   \bottomrule    
        \end{tabular}
    \end{adjustbox}
\end{table*}

Diffusion-based models have been increasingly employed in the field of medical anomaly localization, leveraging various datasets encompassing multiple imaging modalities and anatomical regions. These datasets, summarized in~\cref{tab:datasets}, are critical for developing and validating models that can accurately identify anomalies. Much of the research on diffusion-based anomaly localization methods has centered around brain imaging, notably MR scans. This emphasis is primarily attributed to the abundance of well-annotated datasets and the clinical significance of anomaly localization. Datasets such as BraTS~\citep{brats1,brats2,brats3}, ATLAS~\citep{atlas2022}, and WMH~\citep{WMH} are extensively utilized in this domain, offering multi-modal MRI data with pixel-wise annotations crucial for evaluating anomaly localization models. The expansion into different modalities and anatomies like chest X-rays, retinal optical coherence tomography (OCT) scans, and pediatric wrist X-rays indicates the growing versatility and potential of diffusion-based methods in the broader field of medical imaging. 

However, a significant limitation remains: most datasets traditionally designed for supervised methods focus on a single disease, which is suboptimal for testing the broad detection capabilities of anomaly localization systems. For instance, BraTS is specifically geared towards brain tumors, while ATLAS focuses on stroke detection. These narrow focuses limit the ability to generalize findings across multiple conditions. Additionally, associated non-pathological changes or unrelated diseases, such as mass effects from tumors or atrophy following strokes, are often not annotated. Unsupervised methods can detect these areas as deviations from the norm, identifying them as anomalies, adversely affecting quantitative performance even when the detection is correct.

A few datasets provide multiple labels for different diseases, allowing for a broader spectrum for evaluation. For example, the FastMRI+~\citep{pinaya2022fast,zhao2021fastmri+} dataset includes annotations for 30 different types of anomalies such as tumors, lesions, edemas, enlarged ventricles, resections, post-treatment changes, and more, all marked with bounding boxes. Similarly, the Chexpert~\citep{irvin2019chexpert} dataset, primarily focused on thoracic anomalies, contains annotations for 14 different types of conditions, including pleural effusions, lung opacity, lesions, and cardiomegaly, offering a more comprehensive dataset for evaluation. Additionally, the GRAZPEDWRI-DX (GRAZPED)~\citep{nagy2022pediatric} dataset challenges methods to simultaneously detect and classify various pediatric wrist injuries, including bone anomalies, bone lesions, foreign bodies, fractures, metallic artifacts, periosteal reactions, pronator signs, or soft tissue abnormalities. This diversity in labeling across multiple pathologies enhances the robustness and generalizability of diffusion-based anomaly localization models.

Recent work has begun to explore multimodal extensions to diffusion models in medical imaging. While most current datasets focus solely on image data, some studies \citep{zhang2024diffboost, dong2024diffusion} have incorporated text descriptions to guide generation or segmentation, marking early steps toward true multi-modal pipelines. In addition, \cite{wang2024diffusion} proposed a diffusion-based approach for cross-modality segmentation, demonstrating the potential to leverage complementary information across MR sequences or imaging modalities. This shift toward multimodal learning could enhance model robustness and clinical relevance, making it a promising direction for future research.

In clinical practice, radiologists and domain experts usually specialize in specific organs or imaging modalities, such as neuroradiology or chest imaging. Furthermore, medical imaging datasets are often limited in size, making it challenging to train broadly generalized models effectively. Consequently, adopting a domain-specific focus can be advantageous, as it allows for the development of more accurate models tailored to particular clinical contexts, especially when data are scarce.

As shown in~\cref{tab:datasets}, the number of training samples used for diffusion models varies greatly across studies, reflecting differences in both the task and dataset availability. While some works, such as  \cite{wyatt_anodppm}, evaluated models on as few as 22 MR volumes, others, including \cite{pinaya2022fast}  and \cite{marimont2023disyre}, relied on datasets with over 1,000 volumes. Importantly, the stochastic nature of diffusion training through the progressive addition of noise (see Section 2.1) acts as an inherent and powerful data augmentation mechanism, which can partially compensate for limited data. However, given that medical imaging datasets are typically smaller than those used in general computer vision tasks, the risk of memorization becomes more pronounced. This can lead to overfitting, reduced generalization, and potential privacy concerns \citep{gu2023memorization}. Conditional Denoising Diffusion Probabilistic Models (DDPMs) are significantly more prone to memorization than their unconditional counterparts, because conditioning (e.g., on image features, segmentations, or class labels) introduces informative labels that cause tighter clustering in the model's latent space.  This clustering reduces variance and increases the model’s tendency to reproduce training samples, which is especially problematic in medical imaging, where data scarcity and patient privacy are critical  \citep{chen2024towards}.

\cref{tab:dataset_links} and \cref{tab:github} in the Appendix provide links to publicly available datasets and to GitHub repositories containing the source code.
Furthermore, we refer to the MONAI Generative framework \citep{pinaya2023generative}, which includes a lot of these methods. 
While all of these datasets can be used to develop diffusion-based anomaly localization algorithms, proper comparison and evaluation is crucial to assess the performance and clinical value of the proposed approaches. In~\cref{sec:eval}, we discuss different evaluation metrics for the task of anomaly detection and localization in medical images.

\section{Evaluation Metrics for Anomaly Detection and Localization in Medical Images\label{sec:eval}}
Evaluating the performance of diffusion-based models in medical anomaly detection and localization involves various metrics depending on the available ground truth annotations. 

\paragraph{Image-level Anomaly Detection} For image or volume level, several metrics evaluate a model's ability to classify anomaly presence or absence. The Area Under the Receiver Operating Characteristic Curve (AUROC) is popular due to its threshold-independence and single-value performance summary. However, AUROC may mislead in imbalanced datasets dominated by the majority class. The Area Under the Precision-Recall Curve (AUPRC) and Average Precision (AP) focus on precision-recall trade-offs and offer nuanced evaluations, particularly in imbalanced datasets where anomalies are scarce. The F1 Score, which combines precision and recall into a single metric by calculating their harmonic mean, is useful for balancing false positives and false negatives. However, it requires a specific threshold to be set, which can vary depending on the application, making it sensitive to the balance between precision and recall.
These metrics do not provide spatial information about the detected anomalies but merely indicate whether an anomaly is present, which is a significant limitation when spatial accuracy is critical for clinical applications. \\

\paragraph{Pixel/voxel-wise Localization} is measured with metrics such as pixel-wise AUROC, AUPRC, and the Dice Similarity Coefficient (DSC). The DSC measures the overlap between the predicted and ground truth segmentations, providing a single value that balances false positives and false negatives. However, it requires setting a thresholded anomaly map to determine whether a pixel is classified as anomalous, which can significantly impact results. Selecting an appropriate threshold is challenging and can vary between datasets and applications.
Moreover, these metrics might not be well-calibrated for detecting small anomalies, preferring over-segmentation, where the model identifies larger regions as anomalous than they truly are \citep{maier2024metrics}. Also, they might fail to capture clinically relevant information, such as total lesion load, lesion count, or detection rate.\\ 

\paragraph{Bounding-box Localization} is commonly evaluated with the Intersection over Union (IoU). IoU measures the overlap between the predicted bounding box and the ground truth, with higher IoU indicating better localization. Precision and recall at various IoU thresholds offer a nuanced view of the performance across different levels.
However, generating bounding boxes from continuous anomaly heatmaps produced by diffusion models is challenging. This process can introduce errors and reduce the accuracy of traditional IoU-based metrics. Bercea et al.~\citep{bercea2024generalizing} proposed alternative metrics to address these challenges, such as counting the amount of response inside and outside bounding boxes. However, further development is needed to create metrics that accurately capture clinically relevant localization features.

\section{Types of Supervision for Diffusion-based Anomaly Localization\label{section4}}
\begin{figure*}[t]
    \centering
    \includegraphics[width=0.8\textwidth]{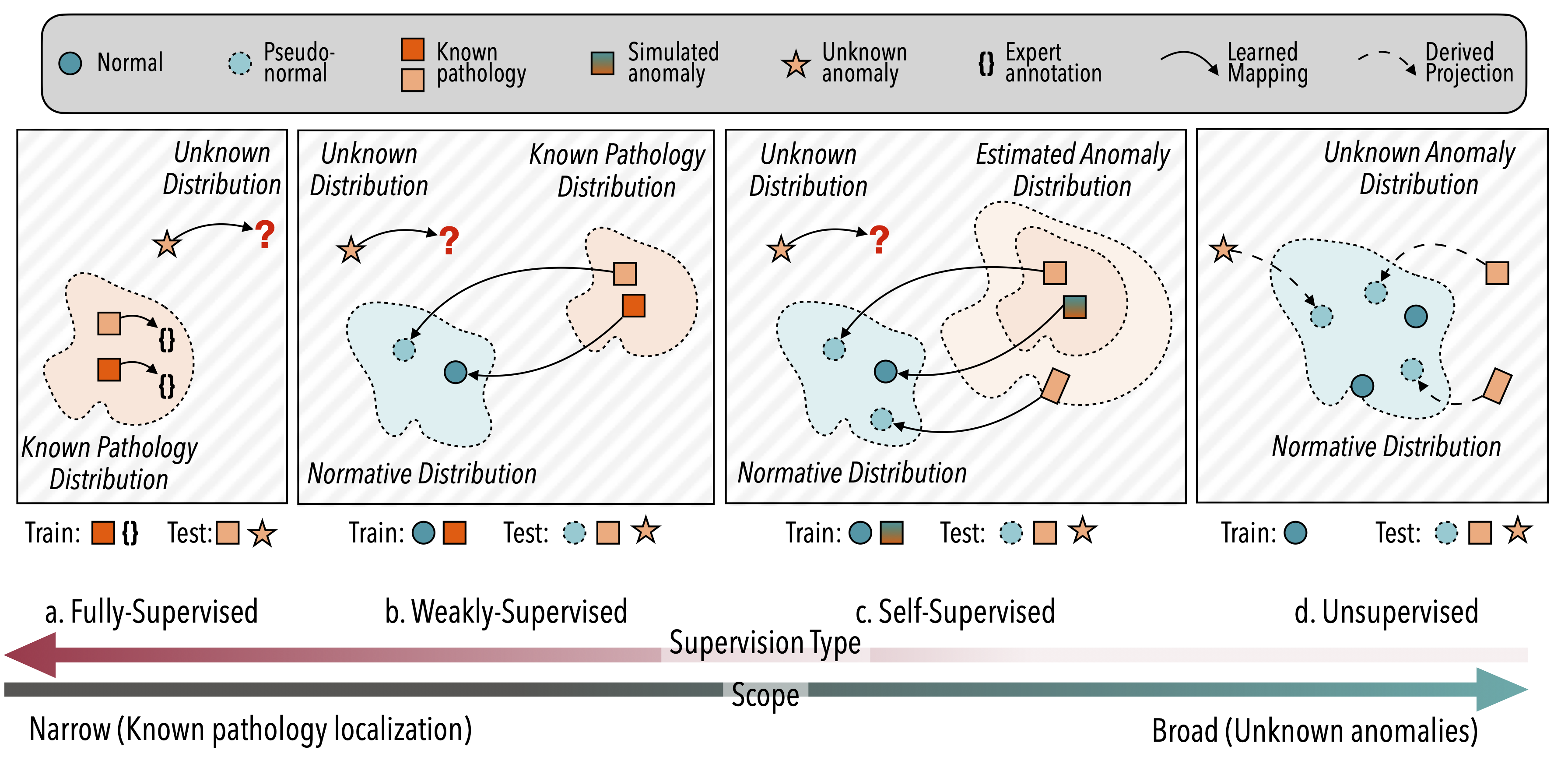}
    \caption{This figure demonstrates the impact of varying levels of supervision on anomaly detection, progressing from targeted pathology localization to the detection of broad, unknown anomalies.
    All symbols are explained in the grey top bar.
    In (a), fully supervised methods use expert annotation to directly predict the pathology segmentation masks. In (b), weakly-supervised methods learn to transform known pathologies (e.g., brain tumors) into pseudo-normal images. In (c), self-supervised methods simulate anomaly types using synthetic augmentations (e.g., coarse noise patterns) that are eliminated to produce pseudo-normal variants. In (a), (b), and (c) behavior outside the learned mappings, such as unknown or rare anomalies depicted as a star, is not defined. In (d), unsupervised methods do not train to learn specific mappings but instead estimate the normative distribution, enabling anomaly detection as deviations from this norm. Semi-supervised methods combine elements of both fully supervised and unsupervised approaches.} 
    \label{fig:supervision}
\end{figure*}
Diffusion models offer a versatile framework for anomaly detection, ranging from targeted pathology localization with supervised models to broad anomaly screening with unsupervised techniques, adapting to the diverse needs of healthcare applications, as shown in~\cref{fig:supervision}. Supervised diffusion models use pixel-wise labels provided by expert clinicians to detect certain pathologies, enabling highly accurate identification and localization of specific anomalies. 
In weakly-supervised settings, which use image-level labels, models transform known pathologies into pseudo-healthy images, allowing for precise localization of these specific anomalies as outlined in~\cref{subsection: basicidea}. Self-supervised models estimate classes of anomaly types through synthetic augmentations. These augmentations, such as coarse noise patterns, are removed to create pseudo-healthy variants, thereby simulating a broader range of anomaly types. However, these models may struggle with unknown or rare anomalies that fall outside the learned mappings.
In these cases, unsupervised approaches that learn a normative distribution are effective in identifying rare or unknown anomalies not present during training due to a lack of information on the expected anomaly distribution.

\subsection{Fully Supervised Lesion Segmentation}

Fully supervised lesion segmentation involves segmenting lesions in medical images using annotated training data, where each pixel is labeled as either belonging to the lesion or not. The training dataset consists of pairs $(i,l)$, with  $i$ being the image and $l$  the corresponding label mask. Diffusion-based approaches use the generative denoising process to produce pixel-wise segmentation masks, leveraging the stochastic nature of diffusion models for output diversity. We want to highlight that fully supervised segmentation approaches directly predict the segmentation mask, and do not rely on the reconstruction pipeline described in~\cref{subsection: basicidea}.

\begin{figure*}[h!]
\centering
\includegraphics[width=0.8\linewidth]{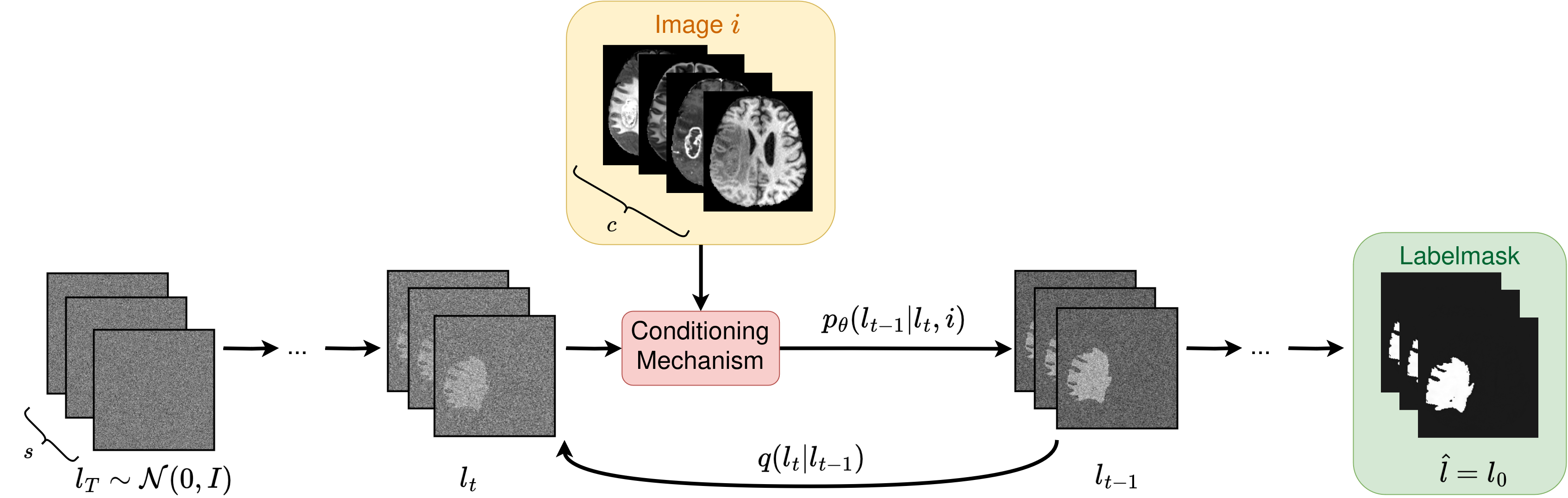}
\caption{The denoising process of DDPMs is leveraged for the generation of a segmentation mask conditioned on the input image $i$. The conditioning mechanism relies on channel-wise concatenation in every step during the denoising process, as described in \cite{saharia2022palette}.}
\label{fig:seg_conditioning}       
\end{figure*}
The technical innovation is presented in~\cref{fig:seg_conditioning}, where the diffusion process is used to generate a segmentation mask. To condition this random generation process on the image to be segmented, we need to incorporate the anatomical information through concatenation. This conditioning technique was first presented in Palette \citep{saharia2022palette}, and then adapted for image segmentation  \citep{wolleb2022segdiff,amit2021segdiff}. This setup enables paired image-to-image translation is represents a major milestone for any medical image-to-image translation task. For this, the technical details are given as follows: 
Starting from random noise $l_T$, the iterative denoising scheme in~\cref{eq:sampling}  is used to obtain a segmentation mask $\hat{l}=l_0$. To generate a label mask for a specific input image $i$, conditioning on the anatomical features of $i$ is required. Various conditioning algorithms have been proposed for DDPMs, one of the first in \textit{Palette} \citep{saharia2022palette}. Given an input image $i \in \mathbb{R}^{c \times h \times w}$, where $c$ is the channel dimension and $h$ and $w$ are the spatial dimensions, the label mask $l$ is of dimension $s \times h \times w$, with $s$ being the number of segmentation classes.

In MedSegDiff \citep{wu2024medsegdiff}, conditioning is achieved using attention modules combined with a feature frequency parser. MedSegDiff-V2 \citep{wu2024medsegdiffv2} extends this by introducing a spectrum-space transformer to learn interactions between semantic features and diffusion noise in the frequency domain, outperforming previous diffusion-based methods.
DermoSegDiff \citep{bozorgpour2023dermosegdiff} leverages diffusion-based segmentation for skin lesions. To improve performance across diverse skin tones and malignancy conditions with limited data, \cite{carrion2023fedd} propose a diffusion-based feature extractor followed by a multilayer perceptron for pixel-wise segmentation.
\cite{zhang2024diffboost} introduces a text- and edge-guided diffusion model that generates diverse, anatomically meaningful synthetic medical images to significantly enhance segmentation performance across multiple imaging modalities.
\cite{tan2023diffss} introduces \textit{DifFSS} a novel few‑shot semantic segmentation paradigm by leveraging diffusion models to generate diverse, intra‑class auxiliary support images conditioned on support masks, scribbles, or boundary maps, thereby enriching support diversity and significantly improving segmentation performance without altering the underlying segmentation network.
\\

\paragraph{Ensembling}
\label{subsec:ensembling}
A key advantage of DDPM-based segmentation is its probabilistic nature, allowing for the sampling of multiple segmentation masks $\hat{l}$ for each input image $i$, resulting in an  ensemble of segmentation masks $\{\hat{l}_1, ... \hat{l}_n\}$ \citep{amit2021segdiff,wolleb2022segdiff}. The pixel-wise variance map of this ensemble highlights image areas where the model predictions are not consistent, serving for model interpretability. This model uncertainty aligns with the variability between human experts, as explored in \citep{rahman2023ambiguous,amit2023annotator}. The role of uncertainty in diffusion-based segmentation is further explored in \cite{zbinden2023stochastic}, where a categorical diffusion model generates multiple label maps accounting for aleatoric uncertainty, i.e., uncertainty originating from inherent randomness or noise in a dataset \citep{hullermeier2021aleatoric}, from divergent ground truth annotations. \\

\paragraph{Going beyond Gaussian Noise}
Diffusion models are not limited to Gaussian noise. While~\cref{fig:seg_conditioning} illustrates the generation process from Gaussian noise along Gaussian trajectories to binary one-hot-encoded masks, this can be improved by using categorical noise. BerDiff \citep{chen2023berdiff} proposes using Bernoulli noise as the diffusion kernel instead of Gaussian noise to enhance binary segmentation. Additionally, \cite{zaman2023surf} suggests a cold diffusion approach for medical image segmentation, involving image perturbations through shifting and horizontal rotation of the segmentation surface. \cite{yan2024cold} replace random noise with segmentation mask conditioning, enhance image features via frequency-domain contrast boosting, and improve focus on target regions using a conditional cross-attention mechanism.\\

\paragraph{Limitations}
While diffusion-based segmentation algorithms perform well in medical lesion segmentation, they suffer from long sampling times compared to classic U-Net approaches. LSegDiff \citep{vu2023lsegdiff} addresses high memory consumption and long sampling times by training a VAE combined with a latent diffusion model.
Additionally, the acquisition and annotation of large, high-quality datasets with pixel-wise lesion annotations is a labor-intensive and time-consuming process. This limits the scope of the model to annotated lesions and introduces potential human bias. Additionally, as described by \cite{wolleb2022segdiff}, there is a higher risk of diffusion-based segmentation to miss small anomalies by predicting empty segmentation masks compared to the classic nnU-Net approach \citep{isensee2021nnu}.
Therefore, other supervision schemes have been explored, as described in the following sections.

\subsection{Semi-Supervised Lesion Segmentation}
\label{sec:3}

\begin{figure*}[t]
\centering
\includegraphics[width=0.9\linewidth]{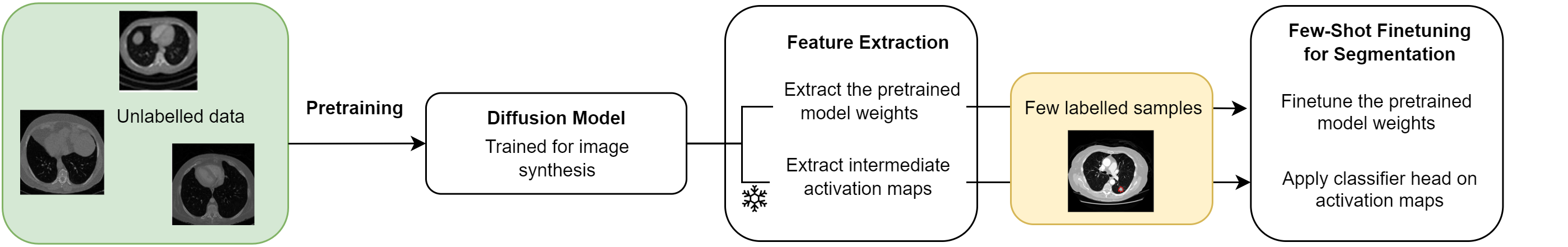}
\caption{Diffusion models excel at distribution learning from unlabeled data. This facilitates effective few-shot fine-tuning on a limited number of labeled samples for lesion segmentation tasks. }
\label{fig:semisupervised}       
\end{figure*}

In semi-supervised learning, algorithms utilize limited labeled data alongside abundant unlabeled data to enhance segmentation accuracy. Typically, pixel-wise labeled data outlining lesion boundaries is scarce and costly to obtain, while large amounts of unlabeled data are often available. Leveraging diffusion models, known for effectively learning data representations, provides an opportunity to learn from this unlabeled data. DDPMs~\citep{baranchuk2021label,yang2023diffusion}, act as representation learners for discriminative computer vision tasks, as shown in~\cref{fig:semisupervised}.
First, a diffusion model is trained for image synthesis on a large unannotated medical dataset to learn a representation of the target anatomy. Feature extraction can then be performed by either extracting the activation maps of the U-Net, or the pretrained model weights. In the final step, fine-tuning with a few pixel-wise labeled images can be done by training a classifier head to predict class labels for each pixel based on the extracted activation maps \citep{alshenoudy2023semi,rosnati2023robust}, or by fine-tuning the extracted model weights on the segmentation task \citep{rousseau2023pre}.
In \cite{behrendt2024combining}, reconstruction-based unsupervised anomaly detection is combined with supervised segmentation on a small annotated dataset to improve segmentation performance for known anomalies and generalization to unknown pathologies.
\cite{ciampi2025semi} present a semi‑supervised teacher‑student framework that uses diffusion models (DDPMs) to generate segmentation pseudo‑labels via unsupervised cycle‑consistency pretraining, and iteratively refines them through multi‑round co‑training to train a student model alongside labeled data. \cite{zhang2024data} introduce a class-conditional diffusion model that generates realistic synthetic image-label pairs by conditioning on segmentation masks, enhancing the diversity of training data.

\subsection{Weakly Supervised Lesion Localization\label{sec:weak}}

\begin{figure}[h]
\centering
\includegraphics[width=\linewidth]{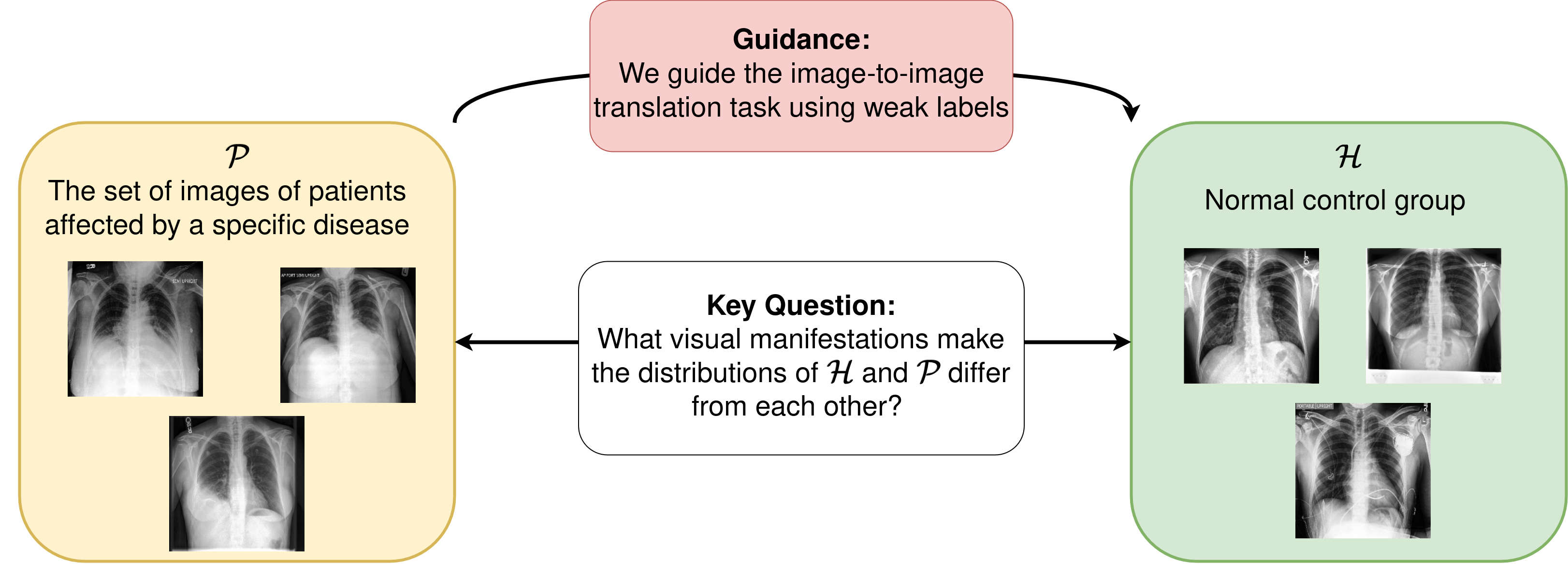}
\caption{In weakly supervised anomaly detection scenarios, two different datasets are at hand. Dataset $\mathcal{H}$ contains images of healthy controls, whereas
dataset $\mathcal{P}$ contains images of patients suffering from a specific disease. Using weakly supervised
methods, the model learns the difference in distribution between $\mathcal{H}$ and $\mathcal{P}$.  }
\label{fig:weaksupervision}       
\end{figure}

Weak supervision involves using imprecise, noisy, or limitedly labeled data to train anomaly detection models. While each image has a label, the annotations lack the detailed information necessary for full supervision.
In weakly supervised anomaly detection, labels are typically derived from image-level information indicating whether the image depicts a patient or a healthy control, as shown in~\cref{fig:weaksupervision}. By utilizing two distinct datasets, $\mathcal{H}$  and $\mathcal{P}$, the objective is to identify the visual manifestations that differentiate between them.
Following an image-to-image translation task, we aim to answer the question  “How would a patient appear if pathology X was not present?” \citep{sanchez2022healthy}. 
A crucial technical milestone was the application of DDIM noise encoding of an input image, as presented in Section 2.2
According to the reconstruction-based method presented in~\cref{subsection: basicidea}, first an input image $x_P$ is encoded to noise using~\cref{eq:ddim_reversed} for $L$ steps. Thereby, all anatomical information is stored in a noisy image $x_L$. This technical innovation enables a loss-free encoding of an input image in noise without destroying information, enabling unpaired image-to-image translation tasks \citep{wolleb2022anomaly,sanchez2022healthy}.
Leveraging a data setup as presented in~\cref{fig:weaksupervision}, weak labels can be used for guidance during denoising.

\paragraph{Gradient Guidance} 
In \cite{wolleb2022anomaly}, a classification network $C$ is trained to distinguish between sets $\mathcal{H}$ and $\mathcal{P}$. Gradient guidance towards the healthy class $\mathcal{H}$ is applied in each denoising step, as described in~\cref{subsec:evolution}. \cite{hu2023conditional} employs a similar approach but locates the desired class by approximating the derivative of the output of a class-conditional diffusion model with respect to the desired label $\mathcal{H}$, rather than using the gradient of an external classification model $C$.
Another gradient-based approach is introduced in \cite{fontanella2023diffusion}, whereby a saliency map from an adversarial counterfactual attention module identifies pathological areas in medical images. During denoising, only these regions are altered using a masking and stitching algorithm.\\

\paragraph{Gradient-free Guidance}
As presented in~\cref{subsec:evolution}, a class-conditional diffusion model $\epsilon(x_t,t,c)$ can be trained on the classes $\mathcal{H}$ and $\mathcal{P}$. 
During denoising, the generation process is guided towards the healthy class by conditioning on $c=\mathcal{H}$ \citep{sanchez2022healthy}.
Building on \cite{ho2022classifier}, in \cite{che2024anofpdm}, the forward process of diffusion models is employed for weakly supervised anomaly detection.
Leveraging class activation maps (CAMs), \cite{yoon2024diffusion} enhance pseudo-mask quality by combining diffusion features with transformer-based CAMs using cross-attention, and enforces patch-level consistency to boost segmentation performance. \\

\begin{figure}[h]
\centering
\includegraphics[width=\linewidth]{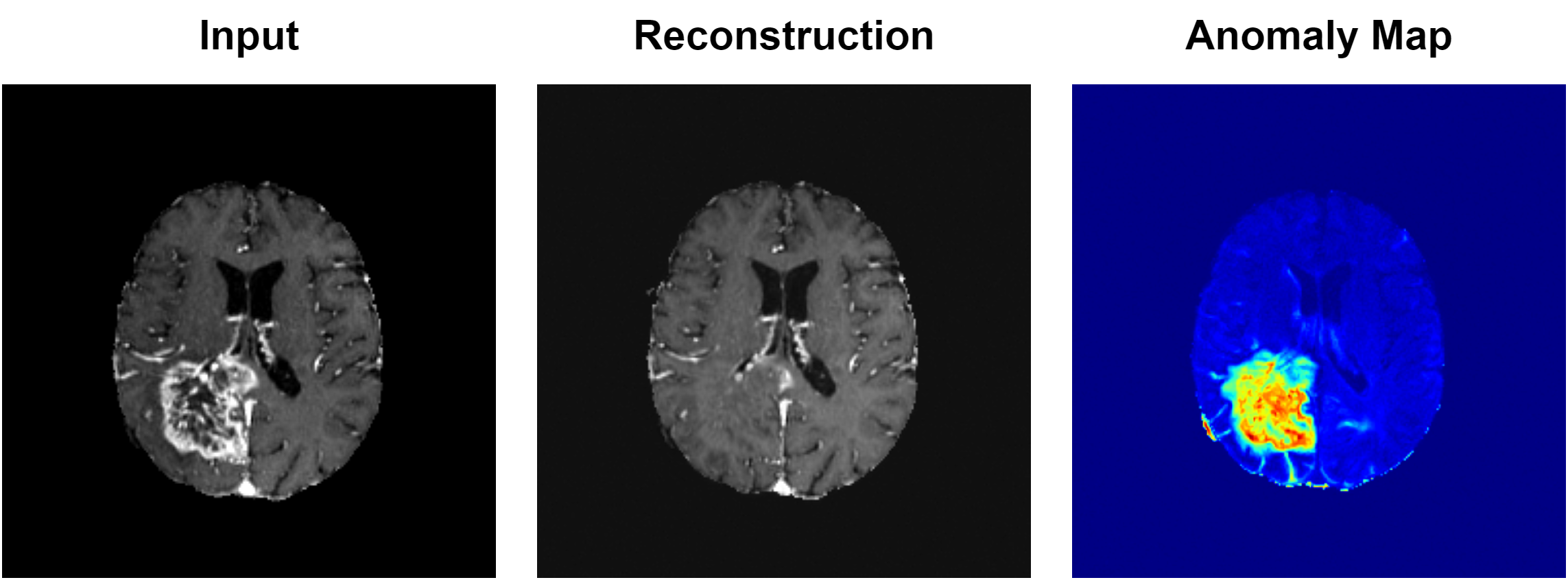}
\caption{Illustration of dataset bias in tumor detection models. The model is trained to distinguish images with and without tumors, successfully removing tumor regions but failing to correct brain deformations caused by large tumors. This results in more accurate difference maps for tumor removal but leaves structural anomalies uncorrected.}
\label{fig:deformation}       
\end{figure}

\paragraph{Limitations}
While weakly supervised methods are more flexible than fully supervised approaches in mimicking human experts' outlining of a specific type of anomaly, they still have limitations in detecting diverse anomalies. Depending on the anomalies present in dataset $\mathcal{P}$, the model is trained to distinguish between the distributions of $\mathcal{P}$ and $\mathcal{H}$, which may result in overlooking other types of anomalies in the input images. 
An example for the dataset bias is shown in~\cref{fig:deformation}. Following \cite{wolleb2022anomaly}, a model is trained to distinguish between images with vs. without tumor. It thereby learns to remove the tumor from the image, but it does not correct from deformations caused by large tumors. On one hand, this renders the difference map more accurate, but also does not correct any anomalies caused by deformation of the brain structure. This specific problem is better handled by VAEs.

\subsection{Self-supervised Anomaly Localization\label{sec:self-sup}}

Self-supervised learning has revolutionized the pre-training of large neural networks, enabling models to learn robust representations from vast amounts of unlabeled data. The core idea behind self-supervised learning is to design auxiliary tasks that do not require manual annotations, allowing the network to learn useful features by solving these tasks. Examples include predicting the rotation angle of an image~\citep{gidaris2018unsupervised} or solving jigsaw puzzles~\citep{noroozi2016unsupervised}.
Recent advancements have demonstrated the efficacy of self-supervised learning in enhancing feature extraction. Self-supervised models can achieve performance on par with, or even surpass, their supervised counterparts in certain tasks. Models like SimCLR~\citep{chen2020simple}, MoCo~\citep{he2020momentum}, and BYOL~\citep{grill2020bootstrap} leverage contrastive learning to maximize agreement between differently augmented views of the same data point, thereby learning powerful and discriminative features. Recently, it has been shown that diffusion models can be self-supervised representation learners, beneficial for many downstream tasks~\citep{xiang2023denoising,yang2023diffusion}.

\begin{figure*}[h]
    \centering
    \includegraphics[width=0.7\linewidth]{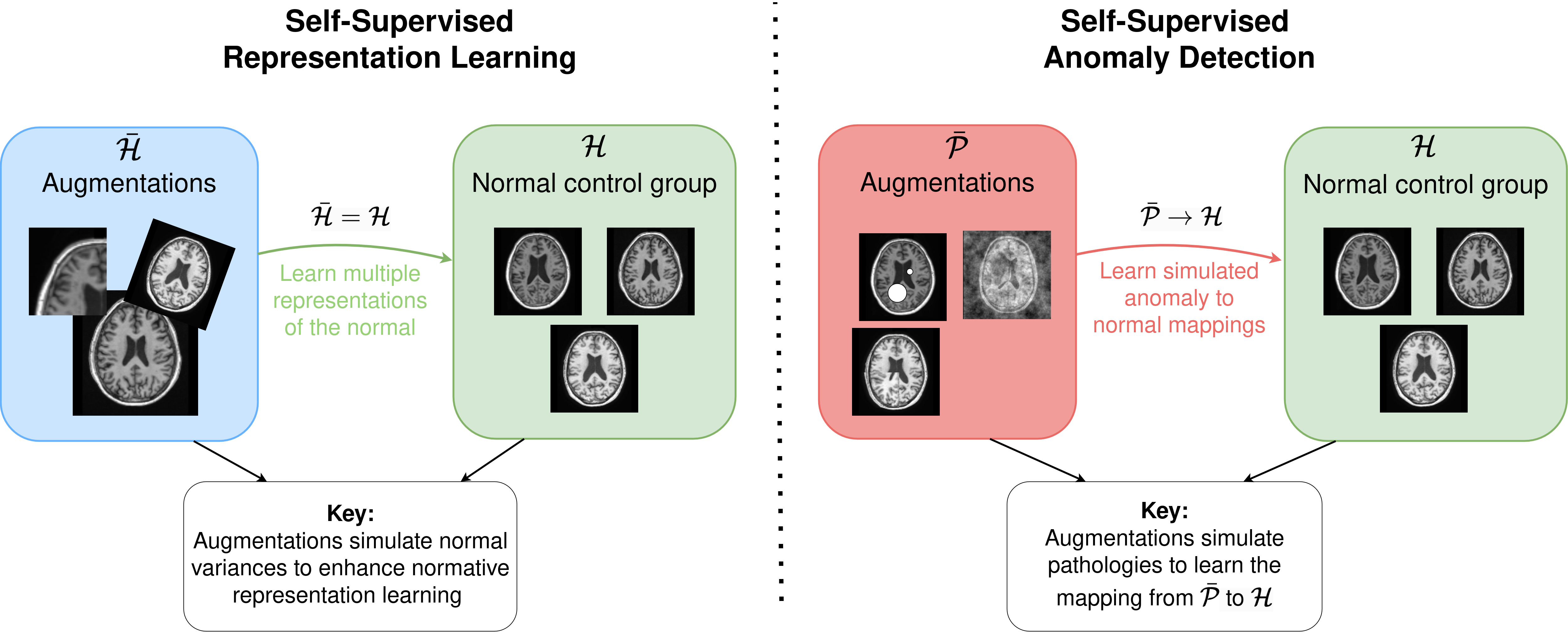}
    \caption{Illustration of self-supervised learning approaches for anomaly detection. Left: Augmentations simulate normal variances to enhance normative representation learning (\(\overline{\mathcal{H}} \approx \mathcal{H}\)). Right: Augmentations simulate pathologies to learn pathology-to-normal mappings (\(\overline{\mathcal{P}} \rightarrow \mathcal{H}\)).}

    \label{fig:self-supervision}
\end{figure*}

In the context of medical imaging, self-supervised pre-training can significantly reduce the dependency on large labeled datasets, which are often difficult and expensive to obtain. By leveraging the inherent structure and properties of medical images, self-supervised methods can learn meaningful representations useful for downstream tasks such as segmentation and localization of lesions. As visualized in~\cref{fig:self-supervision}, self-supervision can be applied in two main ways: learning normative representations and enhancing feature learning, or simulating anomalies in the downstream task. Each approach offers unique advantages and limitations.

\paragraph{Learning Normative Representations} Self-supervised learning can be used to learn representations of healthy data without requiring labeled anomalies. The goal is to simulate a set of augmentations \(\overline{\mathcal{H}}\) of the normative representation and augment the normative training set. Here, the objective is to learn that \(\overline{\mathcal{H}} \approx \mathcal{H}\), thereby capturing a varied distribution within \(\mathcal{H}\). By doing so, the model becomes more adept at identifying deviations from the norm, indicative of anomalies.

Traditionally, this approach has been driven by context encoding VAEs for anomaly detection~\citep{zimmerer2018context}. More recently, transformers that mask parts of the input during training, such as Masked Autoencoders (MAEs)~\citep{he2022masked} and Latent Transformer Models (LTMs)~\citep{pinaya2021unsupervised}, have been used. Lately, this concept was also applied to diffusion-based models by noising only patches from an image and using the rest as context for the denoising process~\citep{behrendt2024patched}.\\ 

\paragraph{Simulating Anomalies}
Another approach involves simulating anomalies as part of the self-supervised learning process. Here, the goal is to simulate pathologies \(\overline{\mathcal{P}}\) and learn the transformation from \(\overline{\mathcal{P}}\) to \(\mathcal{H}\). Various methods have been developed to simulate anomalies. For instance,  one technique involves simulating anomalies by interpolating foreign patches into images and detecting them~\citep{tan2021detecting}. However, such strategies have not yet been adapted to diffusion models. Denoising autoencoders (DAEs) apply coarse noise to a U-Net to simulate anomalies~\citep{kascenas2022denoising}. 

\textit{AnoDDPM}~\citep{wyatt_anodppm} extended this principle by employing diffusion models with \textit{Simplex noise} to simulate structured, spatially coherent perturbations. Unlike Gaussian noise, which acts independently at each pixel, Simplex noise produces continuous patterns across scales, yielding synthetic anomalies that more closely resemble lesion-like structures. A notable aspect of AnoDDPM is its hybrid nature. If a pathology resembles the noise distribution, the denoising process effectively suppresses it---corresponding to a self-supervised form of anomaly removal. However, if the pathology is covered by the noise during inference, the behaviour reduces to the standard unsupervised diffusion setting, where noise is reversed to match the healthy prior patterns. 
\begin{figure*}[t]
    \centering
    \includegraphics[width=0.6\textwidth]{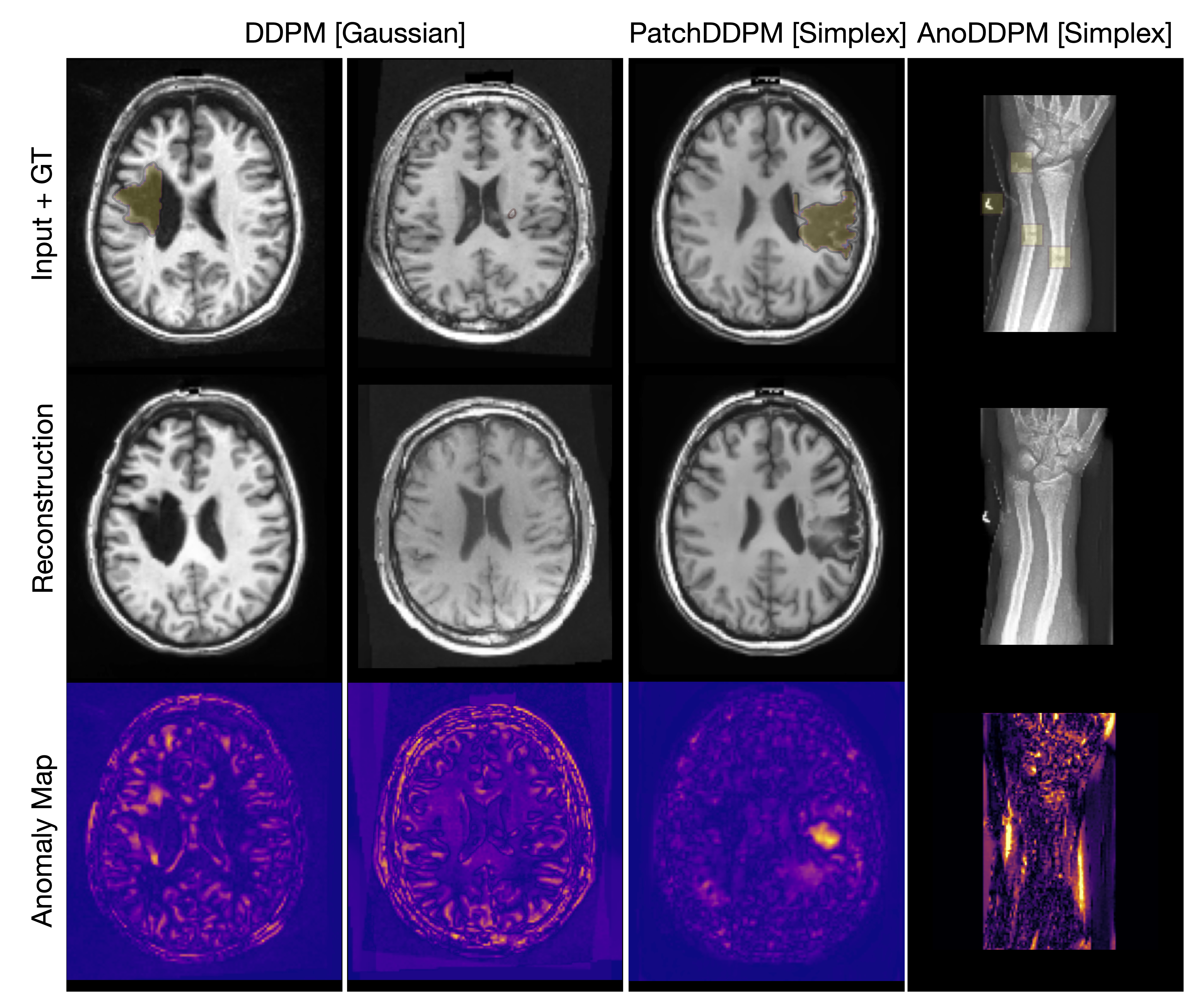}
    \caption{Limitations of diffusion-based anomaly detection in brain and bone MRI, shown from left to right. With Gaussian noise at $L=250$ steps, a stroke lesion next to the lateral ventricle is reconstructed instead of removed, leading to a missed detection. With the same setting, a very small stroke lesion is removed, but the image contains numerous false positives that obscure the result. With PatchDDPM and Simplex noise at 250 steps, only parts of the stroke lesion are removed and the enlarged ventricle is not restored to a normal shape. With AnoDDPM and Simplex noise on bone MRI, noise-like structures such as the fracture are suppressed, yet the reconstruction produces unnaturally bent bones instead of a realistic healthy configuration.}
    \label{fig:diffusion_failures}
\end{figure*}

\paragraph{Limitations}  
Despite their potential, self-supervised anomaly-simulation methods such as AnoDDPM face inherent challenges. Their effectiveness depends on how well the simulated anomalies \(\overline{\mathcal{P}}\) approximate the true pathological variability \(\mathcal{P}\). Since \(\mathcal{P}\) is highly heterogeneous and encompasses many rare diseases\footnote{https://rarediseases.info.nih.gov}, achieving broad anomaly coverage in practice is difficult. Anomalies that diverge from the e.g., Simplex noise distribution may remain undetected or only partially suppressed, as the transformation between real pathologies and their pseudo-healthy reconstructions is not explicitly learned. This limitation is evident in~\cref{fig:diffusion_failures}, where DDPMs with Simplex noise only partially remove stroke lesions and fails to restore enlarged ventricles in brain MRI, or suppresses fracture patterns in bone X-ray but reconstructs them with anatomically-healthy implausible, bent structures. Similar limitations have also been reported by~\cite{bercea2024diffusion,bercea2023reversing}, underscoring the challenge of generalizing beyond the simulated anomaly space.  

\subsection{Unsupervised Lesion Localization}
\label{sec:uad}

Unsupervised anomaly localization methods have emerged as a powerful approach in medical imaging. These methods focus on learning the distribution of normal anatomy (\(\mathcal{H}\)) without any supervision on the expected anomaly distribution, although weak labels of only controls are required to curate a healthy-only dataset. Unlike (self-)supervised methods, unsupervised methods do not aim to learn mappings from pathological sets (\(\mathcal{P}\)) to normal sets (\(\mathcal{H}\)). Instead, they concentrate solely on modeling the distribution of normal, healthy data, as shown in~\cref{fig:supervision}. Deviations from this learned distribution are subsequently classified as anomalies, where pathological images are compared against pseudo-healthy reconstructions (see~\cref{fig:reconstruction}).

Variational autoencoders (VAEs)~\citep{zimmerer2019unsupervised,chen2020unsupervised}  and generative adversarial networks (GANs)~\citep{schlegl2017unsupervised} are commonly used to model the distribution of data of healthy controls. These models are trained to reconstruct healthy images accurately, and any significant deviation in the reconstruction is flagged as a potential anomaly. 

Recently, diffusion models~\citep{ho2020denoising,wyatt_anodppm,liang2023modality} have been applied to this task. An innovative major algorithm was the use of latent diffusion models for unsupervised anomaly localization \citep{pinaya2021unsupervised}. In a first step, an autoencoder based on a VQ-VAE architecture \citep{van2017neural} is trained to encode a healthy image into a low-dimensional latent space. In a second step, a denoising diffusion model is trained in the latent space. During inference, we can evaluate the algorithm on pathological images. The image is passed through the endocer into a latent representation $z$. As this representation will encode unseen changes, the denoising process of the diffusion model, trained for reconstruction of healthy representations, is applied for $L$ steps, to correct for these anomalous changes. The resulting latent vector $\hat{z}$ is then passed through the decoder to reconstruct the pseudo-healthy image. 
However, a main limitation is the loss of healthy tissue information during the reverse process due to the exhaustive noise needed to cover anomalies, referred to as the noise paradox in~\cite{bercea2023mask}. This challenge is illustrated in~\cref{fig:diffusion_failures}, where too little corruption leads to reconstructed anomalies being missed, while excessive corruption removes anomalies but produces false positives. To mitigate this, methods that guide the synthesis to mostly replace only the assumed pathological tissues, using the rest of the healthy context as guidance, have been developed (as discussed in~\cref{subsec:evolution} "Implicit Guidance"). \cite{behrendt2025guided} condition the diffusion model’s denoising process using latent representations of the input image, enabling accurate reconstruction of healthy brain tissue while avoiding reconstruction of anomalies.
\\
\paragraph{Limitations} Despite their potential to detect arbitrary (rare) anomalies without relying on expectations about anomaly distributions, some challenges still remain. Ensuring that the learned distribution of normal anatomy is sufficiently comprehensive to delineate normal variations from subtle anomalies or even detect early alterations of tissues remains a significant challenge. The diversity and complexity of normal anatomical variations require sophisticated modeling techniques to accurately capture these nuances. Moreover, anomaly map computations often rely on pixel-wise differences, which are not ideal for detecting subtle pathological changes due to small intensity differences. Only a few works investigate other ways of computing anomaly maps using perceptual maps~\citep{bercea2024generalizing} or structural similarity~\citep{behrendt2024diffusion}.
Evaluating unsupervised models across diverse datasets and anomaly types remains crucial to ensure their robustness and generalizability.

\section{Open Challenges}
\label{sec:challenges}

While the different types of supervision discussed in~\cref{section4} provide opportunities to train diffusion models for anomaly localization under different data and label availability scenarios, several challenges remain. These include detection bias, high memory requirements for large 3D volumes, distribution shifts in multicentric data, the lack of comprehensive benchmark datasets for clinical validation, computational cost, and model interpretability. In the following subsections, we address these issues in detail.

\subsection{Detection Bias}

Traditionally, anomaly detection methods have been applied to finding multiple sclerosis (MS) lesions or tumors on Fluid-Attenuated Inversion Recovery (FLAIR) brain MR sequences. FLAIR imaging is particularly useful in diagnosing and monitoring conditions like MS, where it excels at revealing hyperintense lesions that indicate areas of demyelination or inflammation. Similarly, it is valuable in identifying other types of brain abnormalities, including tumors and infarctions. This led to an interesting trend where methods that produced blurry reconstructions were found to be more proficient in anomaly localization~\citep{bercea2023aes}. This effect was attributed to the simplicity of the task, which was later shown to be more effectively addressed using simple intensity thresholding techniques~\citep{meissen2021challenging}. Two key lessons can be extracted from this: 

\begin{enumerate}
    \item Anomaly localization should not be evaluated in isolation. Evaluations often only report metrics such as mean absolute error (MAE) or training loss on healthy samples. A more comprehensive approach involves evaluating normative representation learning alongside anomaly localization scores, highlighting the importance of a holistic evaluation framework that provides deeper insights and more meaningful metrics~\citep{bercea2025evaluating}.
    \item Diverse and Comprehensive Datasets: Evaluated datasets should ideally contain a varied array of anomalies, including intensity-based anomalies (e.g., lesions, tumors, inflammation) and structural anomalies (e.g., atrophy, fractures, mass effects). This diversity is crucial to obtaining a comprehensive view of the detection capabilities and limitations of the methods. 
\end{enumerate}
Furthermore, simple baselines and classical literature on anomaly localization should not be overlooked in favor of more advanced generative modeling techniques. Methods should avoid using the vague term "anomaly detection" without clarifying the supervision type, scope, and limitations of their approach. Addressing these challenges necessitates the development of more comprehensive datasets and evaluation metrics that reflect the complexity and variety of medical anomalies. This will enable a more accurate assessment of the true performance and applicability of anomaly detection and localization methods in clinical settings.

\subsection{2D/3D Anomaly Localization}
While almost all approaches presented in~\cref{section4} are implemented in 2D, when dealing with MR and CT scans, 3D approaches will be required. A current challenge are the high memory requirements and long sampling times when implementing 3D diffusion models. To address these issues,  \cite{bieder2023memory} proposed a memory-efficient 3D architecture, enabling a fully supervised lesion segmentation on a resolution of $256^3$ by training only on patches. 
This reduces memory consumption during training from 78.5 GB to 10.6 GB, while maintaining a duration of 1 second for one network evaluation. \cite{durrer2024denoising} further explored architectures for 3D volumes. Running inference on a $256^3$ volume by stacking 2D slices from a 2D model takes around 20 minutes, with memory usage of 34.0 GB. A full 3D approach requires 78.8 GB with roughly the same sampling time. \cite{friedrich2024wdm} demonstrated that 3D wavelet diffusion models can significantly reduce this to 5 minutes per volume with only 32.3 GB of memory.
Another common workaround is the implementation of 3D latent diffusion models \citep{pinaya2022brain,khader2023denoising}, which encode the input data into a compressed latent space before applying the diffusion model. While this approach can be used for anomaly localization \citep{pinaya2022fast,graham2023unsupervised}, model performance is still limited by the lack of a well-performing 3D autoencoder \citep{friedrich2024wdm,durrer2024denoising}. 

Apart from downsampling the 3D volumes, which comes with a loss of information, applying the 2D models slice-wise brings the challenge of missing consistency between the output slices \citep{durrer2024denoising}. This issue is addressed in \cite{zhu2023make}, where pseudo-3D volumes are generated with an additional 1D convolution into the third spatial dimension, ensuring consistent stacks of 2D slices.
\cite{friedrich2024wdm} proposed to apply a discrete wavelet transform to reduce the spatial dimension before applying the diffusion model, enabling processing of volumes of a resolution up to $256^3$. Applications of such architectures on an anomaly localization task still remains to be explored.

\subsection{Non-pathological Distribution Shifts}
A prevalent yet largely unaddressed issue in the development of anomaly localization methods is the occurrence of distributional shifts. Typically, AD methods are trained on a dataset of healthy controls (i.e., show no pathology) and subsequently evaluated on various downstream tasks involving different pathologies. However, these two datasets often originate from different hospitals or scanners, resulting in a notable disparity between the training and evaluation distributions. For instance, routine checks are not commonly conducted in cancer-specific clinics. This discrepancy and other biases can significantly impact anomaly localization performance~\citep{bercea2023bias,meissen2024predictable}.

Some strategies to mitigate this issue involve incorporating slices without clear pathology from a pathological dataset as healthy slices in the training set. However, this approach has drawbacks. These slices may only represent partial brain volumes and might not be entirely healthy, as the non-pathological effects of surrounding lesions are not annotated. Moreover, this practice can lead to data leakage if the same patients are used for both training and evaluation, potentially skewing the results and providing an inaccurate assessment of the effectiveness of the methods. Exploring techniques to adapt to varying distributions during inference is crucial for achieving clinical acceptance and warrants further investigation.

\subsection{Clinical Integration}

To compare performance on benchmark datasets, \cite{bercea2025evaluating} provide a comprehensive evaluation of generative AI models, including diffusion models, for detecting and correcting anomalies in brain MRI scans.
From a clinical perspective, anomaly localization can add value at multiple points in the radiology workflow: (i) case triage and prioritization for suspected urgent findings; (ii) quality control (e.g., motion, missing sequences, laterality mismatches); (iii) decision support within the reporting viewport via anomaly maps and pseudo-healthy reconstructions; (iv) longitudinal change assessment; and (v) safety-net alerts for incidental or rare abnormalities. To be useful, outputs must be calibrated, time-efficient, and straightforward for radiologists to verify.

Key hurdles remain. Beyond retrospective reader studies, prospective, multi-center evaluations are needed to assess effects on reporting time, detection of subtle/rare pathologies, and downstream management. Workflow fit (PACS/RIS integration, turnaround time, human-in-the-loop acceptance/ override), robustness across scanners and protocols, transparency (uncertainty and case-level rationale), and governance (audit trails, versioning, data protection) are prerequisites for regulatory approval and safe deployment.

Existing efforts provide partial progress but are not substitutes for clinical validation. The MOOD dataset~\citep{zimmerer2022mood} evaluates detection of synthetic anomalies on a hidden test set, and the NOVA benchmark~\citep{bercea2025nova} offers evaluation-only data with heterogeneous real cases to probe localization and clinical reasoning. Retrospective analyses exist~\citep{finck2021automated,disorder-free}, and broader method evaluations (including diffusion models) have been reported~\citep{bercea2025evaluating}. What is still lacking are prospective impact studies with predefined clinical endpoints (e.g., reporting time, recall rates, time-to-treatment), standardized reporting checklists, and reference integrations (e.g., open PACS/RIS plug-ins) to lower deployment friction. Until such gaps are addressed, adoption will likely remain limited to research settings or controlled pilots.

\subsection{Computational Considerations}

A significant drawback of denoising diffusion models are the long sampling times due to the iterative generation process, as well as the high memory requirements, mainly due to global attention layers in the U-Net architecture. A large research field has opened to speed up diffusion-based image generation. 
A first step in this direction was already presented in \cite{song2020denoising}. As discussed in~\cref{subsection:theory}, the DDIM denoising schedule can be interpreted as a numerical solver of an ordinary differential equation (ODE). By skipping timesteps during denoising, the step size is increased at the cost of numerical accuracy and image quality \citep{song2020denoising}. Therefore, the design space of appropriate ODE solvers was explored, suggesting Heun's method as a balanced choice between sampling speed and image quality \citep{karras2022elucidating,liu2022pseudo}. Furthermore, faster sampling can be achieved via distillation of the sampling procedure \citep{salimans2022progressive,meng2023distillation}. Another approach is to combine diffusion models with an adversarial component \citep{xiao2021tackling}, or diffusion model sampling can be combined with neural operators \citep{zheng2023fast} for one-step image generation. However, the application of these approaches for anomaly localization tasks still remains to be explored.
\cite{seyfarth2024latent} reveals that  training emissions in 2D latent diffusion models remain relatively stable across image sizes, ranging from $0.40 \pm 0.16$ kg $CO_2$ to $0.79 \pm 0.20$ kg $CO_2$, while 3D models scale sharply with resolution, reaching up to $9.0 \pm 0.9$ kg $CO_2$, equivalent to roughly 92 km of driving. However, synthesis is even more carbon-intensive: generating 10,000 2D samples emits up to $18.2 \pm 0.3$ kg $CO_2$, whereas high-resolution 3D synthesis can emit as much as $362.3 \pm 1.0$ kg $CO_2$, making it the dominant contributor to environmental impact.

\subsection{Interpretability}
Accurate anomaly localization accelerates the diagnostic process and effectively highlights regions of interest. However, interpretability remains a crucial aspect of anomaly localization approaches in medical imaging, particularly for clinical applications. While techniques discussed in~\cref{subsec:ensembling} offer the possibility of estimating model uncertainty pixel-wise, they do not inherently assess the severity or urgency of the findings. Recent advances in large language models (LLMs)~\citep{hua2024medicalclip,zhu2024llms} offer a promising direction for enhancing the interpretability of unsupervised anomaly localization methods. \cite{li2024multi} applied visual question answering models to anomaly detection tasks and demonstrated that LLMs can enhance the interpretability of detected anomalies. Moreover, they showed that anomaly maps used as inputs for LLMs assist them in generalizing to describe unseen anomalies. Nevertheless, more research is needed to fully exploit the potential of these advancements and ensure their effectiveness in clinical practice.

\section{Conclusion}
\label{sec:conclusion}
In conclusion, our exploration of anomaly localization in medical images using diffusion models underscores the nuanced nature of the field. 
Acknowledging that not all approaches are universally effective in all scenarios is crucial. These scenarios encompass varying quantities of available data and corresponding labels, different imaging modalities, and anomaly types. To address these challenges, it is critical to carefully define and tailor evaluation metrics to the specific characteristics of each anomaly type and modality. In addition, exploring alternatives to Gaussian noise and investigating test-time adaptation techniques can help mitigate domain shifts and improve model robustness.

Moreover, the majority of approaches are currently implemented in 2D, highlighting the need for further exploration and adaptation in 3D settings to better capture volumetric anomalies. To this end, fast and memory-efficient diffusion models need to be explored. Recent advances such as flow matching \citep{lipman2022flow,liu2022flow}, shortcut models \citep{frans2024one}, and improved score matching techniques \citep{song2019generative,vahdat2021score} offer promising new directions for the development of diffusion-based generative models. While these methods have been designed primarily for image generation, their potential for downstream applications such as anomaly localization remains largely unexplored, with only very recent work on flow-based, pathology-aware image synthesis \citep{susladkar2025victr} . In this work, in Section 5, we presented various conditioning strategies tailored to different types of denoising diffusion probabilistic or implicit models. However, such techniques will need to be carefully reviewed and adapted to align with the architectural and training differences introduced by these emerging generative paradigms.

Nevertheless, anomaly localization remains a critical challenge in the medical imaging field. Diffusion models offer a promising field of research, particularly through their ability to synthesize high-quality pseudo-healthy images. This capability opens new possibilities for clinical applications, with the potential to enhance diagnostic accuracy and streamline patient care pathways. However, to fully realize their clinical utility, further clinical validation of potential use cases of diffusion-based anomaly localization methods is needed to integrate them into routine medical practice.


\acks{C.I.B. is funded via the EVUK programme ("Next-generation Al for Integrated Diagnostics”) of the Free State of Bavaria and in part supported by Berdelle-Stiftung (grant TimeFlow). Additionally, for the completion of this project C.I.B., was partially supported by the Helmholtz Association under the joint research school 'Munich School for Data Science - MUDS.' J.W. is funded by the Swiss National Science Foundation
(Grant No. P500PT222349), and was  supported by Novartis FreeNovation, and the Uniscientia Foundation.}

%
\ethics{The work follows appropriate ethical standards in conducting research and writing the manuscript, following all applicable laws and regulations regarding treatment of animals or human subjects.}

\coi{We declare we do not have conflicts of interest.}

\data{Datasets discussed in this review are available under the citations and links provided in~\cref{tab:datasets} and \cref{tab:dataset_links}.}

\bibliography{references}
\newpage
\section*{Appendix}
In \cref{tab:dataset_links} we provide the links to publicly available datasets reported in \cref{tab:datasets}. In \cref{tab:github}, we provide links to the Github code repositories.

\begin{table*}[h] 

    \centering
    \caption{Links to publicly available datasets used in diffusion-based medical anomaly detection.\label{tab:dataset_links}}
    \begin{tabular}{l | l} 
    \toprule
    Dataset & Link \\ \midrule
    BraTS & https://www.med.upenn.edu/cbica/brats2020/data.html \\
    ATLAS & https://atlas.grand-challenge.org/ \\
    WMH & https://wmh.isi.uu.nl/\\
    MSLub & https://pubmed.ncbi.nlm.nih.gov/29995847/ \\
    FastMRI+ & https://arxiv.org/abs/2109.03812\\
    CheXpert & {https://stanfordmlgroup.github.io/competitions/chexpert/} \\
    OCT2017 & https://data.mendeley.com/datasets/rscbjbr9sj/2 \\
    GRAZPED & https://figshare.com/articles/dataset/GRAZPEDWRI-DX/14825193 \\
    \bottomrule
    \end{tabular}
\end{table*}

\begin{table*}[h!] 
\small
\centering
\caption{Key diffusion-based anomaly detection and segmentation methods with direct links. \label{tab:github}}
\renewcommand{\arraystretch}{1.2}
\setlength{\tabcolsep}{8pt}
\begin{tabular}{l l} 
\toprule
Method & First author \\
\midrule
\multicolumn{2}{l}{\textbf{Fully supervised lesion segmentation}} \\
SegDiff & Tomer Amit \\
\quad$\rightarrow$ \textit{https://github.com/tomeramit/SegDiff} & \\
Diffusion-based Segmentation Ensembles & Julia Wolleb \\
\quad$\rightarrow$ \textit{https://github.com/JuliaWolleb/Diffusion-based-Segmentation} & \\
MedSegDiff & Junde Wu \\
\quad$\rightarrow$ \textit{https://github.com/SuperMedIntel/MedSegDiff} & \\
DermoSegDiff & Afshin Bozorgpour \\
\quad$\rightarrow$ \textit{https://github.com/xmindflow/DermoSegDiff} & \\
BerDiff (Conditional Bernoulli Diffusion) & Tao Chen \\
\quad$\rightarrow$ \textit{https://github.com/takimailto/BerDiff} & \\
\midrule
\multicolumn{2}{l}{\textbf{Self-supervised anomaly simulation}} \\
AnoDDPM (Simplex noise) & Julian Wyatt \\
\quad$\rightarrow$ \textit{https://github.com/Julian-Wyatt/AnoDDPM} & \\
PatchDDPM (Patched Diffusion Models for UAD) & Finn Behrendt \\
\quad$\rightarrow$ \textit{https://github.com/FinnBehrendt/patched-Diffusion-Models-UAD} & \\
DISYRE (Diffusion-Inspired Synthetic Restoration) & Sergio N. Marimont \\
\quad$\rightarrow$ \textit{https://github.com/snavalm/disyre} & \\

\midrule
\multicolumn{2}{l}{\textbf{Weakly supervised lesion localization}} \\
Diffusion Models for Medical Anomaly Detection & Julia Wolleb \\
\quad$\rightarrow$ \textit{https://github.com/JuliaWolleb/diffusion-anomaly} & \\
What is Healthy? Generative Counterfactual Diffusion & Pedro Sanchez \\
\quad$\rightarrow$ \textit{https://github.com/vios-s/Diff-SCM} & \\
AnoFPDM & Yiming Che \\
\quad$\rightarrow$ \textit{https://github.com/SoloChe/AnoFPDM} & \\
\midrule
\multicolumn{2}{l}{\textbf{Unsupervised lesion localization}} \\
AutoDDPM (Mask--Stitch--Re-Sample) & Cosmin I. Bercea \\
\quad$\rightarrow$ \textit{https://github.com/ci-ber/autoDDPM} & \\
Fast UAD with Diffusion Models & Walter H. L. Pinaya \\
\quad$\rightarrow$ \textit{https://github.com/Project-MONAI/GenerativeModels/tree/main/tutorials/generative/2d\textunderscore ldm} & \\

cDDPM (Conditioned DDPMs for UAD) & Finn Behrendt \\
\quad$\rightarrow$ \textit{https://github.com/FinnBehrendt/Conditioned-Diffusion-Models-UAD} & \\

MMCCD (Modality Cycles + Masked Conditional Diffusion) & Ziyun Liang \\
\quad$\rightarrow$ \textit{https://github.com/ZiyunLiang/MMCCD} & \\
Binary-noise UAD (Anomaly\_berdiff) & Julia Wolleb \\
\quad$\rightarrow$ \textit{https://github.com/JuliaWolleb/Anomaly\textunderscore berdiff} & \\
THOR (Temporal Harmonization for Optimal Restoration) & Cosmin I. Bercea \\
\quad$\rightarrow$ \textit{https://github.com/ci-ber/THOR\textunderscore DDPM} & \\
\bottomrule
\end{tabular}
\end{table*}


\end{document}